\documentclass[epj]{svjour}
\usepackage{url}
\usepackage{hyperref}
\usepackage{xspace}
\usepackage{graphicx}
\usepackage{amssymb}
\usepackage{multirow}
\usepackage{array}
\usepackage[usenames,dvipsnames]{color}
\usepackage[pagewise]{lineno}

\usepackage{graphics}
\newcommand{\sqsn}{\mbox{$\sqrt{s_{_{NN}}}$}\xspace}

\begin{document}
\title{Multiplicity fluctuations in heavy ion collisions using canonical and grand canonical ensemble}
\author{\large P.~Garg$^{1,}$\thanks {e-mail: prakhar@rcf.rhic.bnl.gov}, \large D.~K.~Mishra$^{2,}$\thanks {e-mail: dkmishra@rcf.rhic.bnl.gov}, \large P.~K.~Netrakanti$^{2}$ \and \large A.~K.~Mohanty$^{2}$
}                     
%
%
\institute{Discipline of Physics, School of Basic Science, Indian Institute of Technology 
Indore 452017, India \and Nuclear Physics Division, Bhabha Atomic Research Center, Mumbai 400085, India}
\date{Received: date / Revised version: date}
%
\abstract{
We report the higher order cumulants and their ratios for baryon, charge and strangeness 
multiplicity in canonical and grand-canonical ensembles in ideal thermal model including 
all the resonances. When the number of conserved quanta is small, an explicit treatment of 
these conserved charges is required, which leads to a canonical description of the system and 
the fluctuations are significantly different from the grand canonical ensemble. Cumulant ratios 
of total charge and net-charge multiplicity as a function of collision energies are also
compared in grand canonical ensemble.%
} 
\authorrunning{P. Garg et al.}
\maketitle
\section{Introduction}
\label{intro}
Event-by-event fluctuations of particle multiplicities and transverse energy have been of great
interest and are measured at AGS, SPS, RHIC and LHC energies in relativistic heavy-ion 
collision experiments \cite{Heiselberg:2000fk,Alner:1985zc,Alner:1987wb,Appelshauser:1999ft,Abelev:2008jg,Adamczyk:2013dal,Adare:2008ns,Abelev:2012pv}. The main motivation of these 
studies is to explore the phase transition and/or a critical end point (CEP), which is believed 
to exist somewhere between the hadronic phase and the quark-gluon phase of the QCD phase 
diagram \cite{Gavai:2010zn,Bazavov:2012vg,Stephanov:2011pb,Asakawa:2009aj,Gazdzicki:2003bb}. 
Most commonly measured event-by-event fluctuations in heavy ion collision experiments are 
particle ratios ($K/\pi$, $p/\pi$ etc.), transverse energy ($\langle E_{T} \rangle$), 
transverse momentum ($\langle p_{T} \rangle$) and multiplicity ($\langle N \rangle$) 
fluctuations \cite{Abelev:2009ai,Alt:2008ab,Adcox:2002pa,Adler:2003xq}. Recently, the higher 
moments of conserved numbers measured in beam energy scan (BES) program at RHIC have 
attracted further attention towards the usefulness of fluctuation studies in heavy-ion 
collisions ~\cite{Adamczyk:2013dal,Adare:2015aqk,Adamczyk:2014fia}. In addition, it has been 
proposed that the kurtosis of the order parameter becomes negative when the critical point is 
approached and as a result, the kurtosis of a fluctuating observable ~\cite{Stephanov:2011pb}, 
e.g., proton multiplicity, may become smaller than the value given by independent Poisson 
statistics. In the present work we explore the fact that, the entropy is closely related to the 
particle multiplicity, and it is expected to be approximately conserved during the evolution of 
the matter created at the early stage of the collision. Therefore, the higher order 
fluctuations of entropy would also be an interesting observable to look for the possibility of 
phase transition and critical point \cite{Gazdzicki:2005mx,Heiselberg:2000ti}. The entropy 
fluctuations are not directly observed but can be inferred from the experimentally measured 
quantities. The system's entropy is related to the mean particle multiplicity, as the final 
state mean multiplicity is proportional to the entropy of the initial state ($\langle N \rangle 
\sim S$) \cite{Jeon:2003gk}. The particle multiplicity can be measured on an event-by-event 
basis, whereas the entropy is defined by averaging the particle multiplicities in the ensemble 
of events. Thus, the dynamical entropy fluctuations can be measured experimentally by measuring 
the fluctuations in the mean multiplicity. Recently measured higher order fluctuations of 
net-proton multiplicity distributions shows that, at lower energies net-proton fluctuation is 
mostly dominated by proton, as the anti-proton production is very 
small~\cite{Adamczyk:2013dal}. Hence, measuring only proton fluctuation will also provide the 
similar conclusion as measuring net-proton fluctuation. 

Assuming a thermal system, formed in heavy ion collisions, the grand canonical ensemble (GCE) 
is the most appropriate description as only part of the particles from the system around 
mid-rapidity are measured by experiments. In heavy-ion collision, if one could make the 
measurements with full phase-space coverage, no conserved number fluctuation would be seen, as 
baryon number ($B$), electric charge ($Q$) and strangeness ($S$) are strictly conserved. In 
thermal models, the magnitude of multiplicity fluctuations and correlations in limited 
phase-space crucially depends on the choice of the statistical ensemble that imposes different 
conservation laws~\cite{Hauer:2009nv}. The micro canonical ensemble (MCE) considers all the 
micro states where energy, momentum and charge are conserved. The canonical ensemble (CE) 
relaxes the energy conservation by introducing an infinite heat bath which can exchange energy 
but conserves the charge. The GCE introduces chemical potential and the requirement of charge 
conservation is also dropped. The multiplicity fluctuation patterns in full or finite momentum 
space are very different in MCE and CE as both impose different conservation conditions 
\cite{Begun:2004pk,Begun:2004zb,Hauer:2007im}. The fluctuations of the energy ($\langle E 
\rangle$) are identical between CE and GCE, but fluctuations of particles which carry the 
conserved charge are affected. Since in GCE, the energy and conserved numbers may be exchanged 
with the rest of the system, therefore, they may fluctuate on an event-by-event basis. The 
experimentally measured multiplicity and transverse energy fluctuations can be related to the 
number susceptibilities and the heat capacity of the system, respectively \cite{Jeon:2003gk}. 

If the number of conserved quanta is small, the grand canonical approach is not 
adequate~\cite{Jeon:2003gk}. Instead the description needs to ensure that the quantum number is 
conserved explicitly in each event. The CE has been used to describes the system formed in 
$p+p(\bar{p})$ and $e^{+}+e^{-}$ collisions where the particle production is small 
\cite{Becattini:1996}. In such systems, the deposited energy is still large and distributed 
over many degrees of freedom, hence the canonical treatment is the appropriate ensemble. 
Further, at lower energies, due to production threshold for strange particles or anti-baryons, 
one can apply the CE prescription to describe system formed in the heavy-ion collisions. 
Fluctuation results from CERN SPS with lower energies and different collision 
species~\cite{Stefanek:2014pza} motivate us to study the fluctuation variables in CE and 
compare the results obtained from those in GCE.

The paper is organized as follows: Section~\ref{sec:ce_gce} describes the formalism for CE 
and GCE partition functions and the cumulants for the total charge fluctuations. In 
Section~\ref{sec:results}, the comparison of cumulant ratios for baryons, charge and 
strangeness number obtained in CE and GCE are discussed. Comparison of total charge and 
net-charge multiplicity fluctuations using GCE are discussed in Section~\ref{sec:totalnet}. We 
summarize the present work in Section~\ref{sec:summary}. 

\section{Canonical and Grand canonical partition functions and their corresponding cumulants}
\label{sec:ce_gce}
Let us consider a system of particles and their corresponding anti-particles. In Boltzmann 
approximation, the grand canonical partition function can be written as \cite{Begun:2004zb}:
\begin{eqnarray}
\label{Zgce}
Z_{gce}(V,T,\mu)
& =&\sum_{N_{1+},N_{1-}=0}^{\infty}\ldots\sum_{N_{j+},N_{j-}=0}^{\infty}
\frac{\left(\lambda_{1+}z_1\right)^{N_{1+}}}{N_{1+}!}   \nonumber \\
&&\frac{\left(\lambda_{1-}z_1\right)^{N_{1-}}}{N_{1-}!}\ldots
\frac{\left(\lambda_{j+}z_j\right)^{N_{j+}}}{N_{j+}!}
\frac{\left(\lambda_{j-}z_j\right)^{N_{j-}}}{N_{j-}!}\ldots \nonumber \\
&=& \prod_{j}\sum_{N_{j+},~N_{j-}=0}^{\infty}
\frac{\left(\lambda_{j+}z_j\right)^{N_{j+}}}{N_{j+}!}
\frac{\left(\lambda_{j-}z_j\right)^{N_{j-}}}{N_{j-}!} \nonumber \\ 
&=& \prod_j\exp\left(\lambda_{j+}z_j+\lambda_{j-}z_j\right)  \nonumber \\
&=&\exp\left[2z\cosh\left(\frac{\mu}{T}\right)\right]~, 
\end{eqnarray}

Here $\lambda_{j\pm}$ = $\mathrm {exp}(\pm \mu/T)$ corresponds to the fugacity of $j^{th}$ 
particle and $\mu$ is the chemical potential. The $"+"$ and $"-"$ signs correspond to the particle 
and anti particle, respectively. And, $z\equiv \sum_{j}z_j$, where $z_j$ is the single particle partition 
function defined as follows;
\begin{eqnarray}
\label{z}
z_j &=& \frac{g_jV}{2\pi^2}
       \int_{0}^{\infty}p^{2} dp\;
       \exp\left[-~\frac{(p^{2}+m_j^{2})^{1/2}}{T}\right] \nonumber \\
        &=& \frac{g_jV}{2\pi^2}  
       T\,m_j^2\,K_2\left(\frac{m_j}{T}\right),
\end{eqnarray}
here $m_j$ is the mass of a $j$-th particle, $K_2$ is the modified Hankel function, 
$T$ and $V$ are temperature and volume of the system, respectively. 

In canonical ensemble, the number of particles are strictly conserved and only the energy can be 
exchanged with the system's surrounding, hence the chemical potential is zero which leads to charge 
conservation constraint $\langle Q\rangle = \langle N_{+} \rangle - \langle N_{-}\rangle = 0$ 
and the partition function reads as follows 
\cite{Begun:2004zb,Begun:2004gs}:
\begin{eqnarray}
\label{Zce}
Z_{ce}(V,T)
 &=& \sum_{N_+=0}^{\infty}\sum_{N_-=0}^{\infty}\;
 \frac{(\lambda_+ z)^{N_+}}{N_+!}\;\frac{(\lambda_- z)^{N_-}}{N_-!}
 \;\delta (N_+-N_-) \nonumber
 \\
 &=&
 \frac{1}{2\pi}\int_0^{2\pi}d\phi\;\;
   \exp\left[ z\;(\lambda_+\;e^{i\phi} 
                   \;+\; \lambda_-\;e^{-i\phi})\right] \nonumber \\
  &=& I_0(2z)\;. 
\end{eqnarray}
Further, the CE  partition function can be modified for an explicit charge conservation 
constrain, i.e. $\sum_{j}\left(N_{j+} - N_{j-}\right) = Q\;$, for each microscopic state 
of the system \cite{Begun:2004zb}:
\begin{eqnarray}
\label{ZceQ}
\lefteqn{ Z_{ce}(V,T,Q) =} \nonumber \\
&&\sum_{N_{1+},~N_{1-}=0}^{\infty}...\sum_{N_{j+},~N_{j-}=0}^{\infty} \nonumber
 \frac{\left(\lambda_{1+}z_1\right)^{N_{1+}}}{N_{1+}!}~ \\
&& \frac{\left(\lambda_{1-}z_1\right)^{N_{1-}}}{N_{1-}!}~... \frac{\left(\lambda_{j+}z_j\right)^{N_{j+}}}{N_{j+}!}  \frac{\left(\lambda_{j-}z_j\right)^{N_{j-}}}{N_{j-}!}...   \nonumber \\
&&\times \delta\left[\left(N_{1+}+...+N_{j+}+... -N_{1-}-...-N_{j-}-...\right) -Q\right] \label{eq:zceq}  \nonumber\\
\\
&=&\int_0^{2\pi}\frac{d\phi}{2\pi}~ \prod_{j}
\sum_{N_{j+},N_{j-}=0}^{\infty}\; 
 \frac{\left(\lambda_{j+} z_j\right)^{N_{j+}}}{N_{j+}!}
\frac{\left(\lambda_{j-} z_j\right)^{N_{j-}}}{N_{j-}!}  \nonumber \\
&&\times \exp\left[i\left(N_{j+}-N_{j-}-Q\right)\phi\right]  \nonumber \\
&=& \int_0^{2\pi}\frac{d\phi}{2\pi}\;\;
\exp\left[-i\,Q\,\phi \;+\;\sum_{j} z_j\;\left(\lambda_{j+}\;e^{i\phi}
  \;+\; \lambda_{j-}\;e^{-i\phi}\right)\right] \nonumber \\
&=&I_Q(2z).
\end{eqnarray} 

In equation (\ref{eq:zceq}), the integral representations of the $\delta$-Kronecker symbol and 
the modified Bessel function are defined as~\cite{Abramowitz}:
\begin{eqnarray}
\label{IQ}
\delta(n) = \frac{1}{2\pi}\int_0^{2\pi}d\phi~ \exp(in\phi)~,\nonumber \\
I_Q(2z) =  \frac{1}{2\pi}\int_0^{2\pi}d\phi \exp[-i Q \phi+ 2z\cos\phi] .
   \end{eqnarray}
It is to be noted that in equation (\ref{eq:zceq}), $\lambda_{j+}$ and $\lambda_{j-}$ are not fugacities but just auxiliary parameters, only to calculate the mean number and the fluctuations of positively and negatively charged particles. They are set to one in the final formula.  

Using the above partition functions for CE, one can derive the other thermodynamic properties 
of the system at freeze-out. Commonly, the mean multiplicity and variance of the particle number distributions are derived 
using the partition functions. The cumulants of multiplicity distribution in GCE can be derived 
as follows:
\begin{eqnarray}
\label{gce_mean}
\langle N_{\pm}\rangle_{gce}
& = &
 \left( \frac{\partial}{\partial\lambda_{\pm}}\ln Z_{gce}
 \right) =\; \lambda_{\pm} z ~,\\
\langle N^{2}_{\pm}\rangle_{gce}
& = &
\frac{1}{Z_{gce}} \left(  \lambda_{\pm} \frac{\partial}
{\partial\lambda_{\pm}} \right )^{2} Z_{gce}
=\; z\lambda_{\pm} +z^{2}\lambda^{2}_{\pm}
\end{eqnarray}

In the present work, we have extended these studies to the higher order cumulants. Hence, 
third and fourth cumulants of the particle and anti-particle multiplicities are derived as, 
\begin{eqnarray}
\label{gce_m3}
\langle N^{3}_{\pm}\rangle_{gce}
& = &
\frac{1}{Z_{gce}}\left( 
\lambda_{\pm}\frac{\partial}{\partial\lambda_{\pm}}\right)^{3} Z_{gce} \nonumber \\
&=& z\lambda_{\pm} +3z^{2}\lambda^{2}_{\pm}+z^{3}\lambda^{3}_{\pm} ~,\\
\langle N^{4}_{\pm}\rangle_{gce}
& = &
\frac{1}{Z_{gce}}\left( 
\lambda_{\pm}\frac{\partial}{\partial\lambda_{\pm}}\right)^{4} Z_{gce} \nonumber \\
 &=&  z\lambda_{\pm} +7z^{2}\lambda^{2}_{\pm}+6z^{3}\lambda^{3}_{\pm} ~.
\end{eqnarray}

Similarly, in CE, the cumulants can be easily derived using the CE partition 
function as defined in equation (\ref{ZceQ}),
\begin{eqnarray} 
\label{ce-ave}
\langle N_{\pm}\rangle_{ce}
& = &
z \frac{I_{Q\mp1}(2z)}{I_{Q}(2z)} ~,\nonumber \\
\langle N_{\pm}^{2}\rangle_{ce} 
&=& 
z \frac{I_{Q\mp1}(2z)}{I_{Q}(2z)} +z^{2} \frac{I_{Q\mp2}(2z)}{I_{Q}(2z)} ~, \nonumber \\
\langle N^{3}_{\pm}\rangle_{ce}
& = &
z \frac{I_{Q\mp1}(2z)}{I_{Q}(2z)} +3z^{2} \frac{I_{Q\mp2}(2z)}{I_{Q}(2z)} \nonumber \\
&+& z^{3} \frac{I_{Q\mp3}(2z)}{I_{Q}(2z)} ~,\\
\langle N^{4}_{\pm}\rangle_{ce}
& = &
z \frac{I_{Q\mp1}(2z)}{I_{Q}(2z)} +7z^{2} \frac{I_{Q\mp2}(2z)}{I_{Q}(2z)} \nonumber \\
&+& 
6z^{3} \frac{I_{Q\mp3}(2z)}{I_{Q}(2z)}+z^{4} \frac{I_{Q\mp4}(2z)}{I_{Q}(2z)} ~,
\end{eqnarray}
and the correlation between particles and their anti-particles can be estimated using the 
following generalized relation:
\begin{eqnarray} 
\label{gce-average}
\langle N^{n_{1}}_{\pm}N^{n_{2}}_{\mp}\rangle \;=\;
\frac{1}{Z}\left( 
\lambda_{\pm}\frac{\partial}{\partial\lambda_{\pm}}\right)^{n_{1}}
\left(\lambda_{\mp} \frac{\partial}{\partial\lambda_{\mp}} \right)^{n_{2}} Z
\end{eqnarray} 
Using the above relations cumulants of the charge multiplicity in both GCE and CE can 
be obtained as follows:
\begin{eqnarray}    
\label{eq:cn_1}
C_{1}
& = & \langle N_{+}+N_{-}\rangle\;=\langle N_{+}\rangle+\langle N_{-}\rangle 
~,\\
\label{eq:cn_2}
C_{2}
& = &\langle (\delta{N})^{2}\rangle\;=\;
\langle (N_{+} + N_{-})^{2}\rangle   \nonumber \\ 
&-&\; \langle (N_{+} + N_{-})\rangle^{2} ~,\\
\label{eq:cn_3}
C_{3}
& = &\langle (\delta{N})^{3}\rangle =
\langle (N_{+} + N_{-})^{3}\rangle \nonumber \\
&-&3 \langle (N_{+} + N_{-})^{2}\rangle\langle (N_{+} + N_{-})\rangle \nonumber \\
&+&2\;\langle (N_{+} + N_{-})\rangle^{3} ~,\\
\label{eq:cn_4}
C_{4}
& = &\langle (\delta{N})^{4}\rangle - 3 \langle (\delta{N})^{2}\rangle ^2
\nonumber \\
& = &
\langle (N_{+} + N_{-})^{4}\rangle -4 \langle (N_{+} + N_{-})^{3} \rangle 
\;C_{1} \nonumber \\
&+&
6\langle (N_{+} + N_{-})^{2}\rangle \; C_{1}^{2} - \; 3C_{1}^{4} -\; 
3C_{2}^{2} ~.
\end{eqnarray}

The properties of distribution functions are characterized by the various 
moments, such as mean ($M$), variance ($\sigma$), skewness ($S$) and kurtosis ($\kappa$). 
These moments are the alternative methods to characterize a distribution besides the 
cumulants. Various moments and cumulants are related as:
$M=C_{1}$, $\sigma^{2}=C_{2}$, $S=C_{3}/C_{2}^{3/2}$ and $\kappa=C_{4}/C_{2}^{2};$
and hence their ratios and products can be written in term of cumulants as:
$\sigma^{2}/M = C_{2}/C_{1}$ , $S\sigma= C_{3}/C_{2}$ and $\kappa\sigma^{2} =C_{4}/C_{2}$. 
Experimentally, one measures the multiplicity distributions of particles (both $N_{+}$ and 
$N_{-}$) on an event-by-event basis and construct the  ($N_{+} + N_{-}$) for total and 
($N_{+} - N_{-}$) for net-charge multiplicity distribution. Recently, net-baryon (proton), 
net-electric charge and net-strangeness (kaon) fluctuations measured in BES at RHIC have 
further attracted attention towards the event-by-event fluctuation studies using their higher 
moments ~\cite{Adamczyk:2013dal,Adare:2015aqk,Adamczyk:2014fia}. Ratios and products of the 
moments of total multiplicity distributions can also be experimentally measured and it will be 
interesting to see their dependences on the collision energy (\sqsn). 
\begin{figure}[h]
 \begin{center}
\includegraphics[scale=0.43]{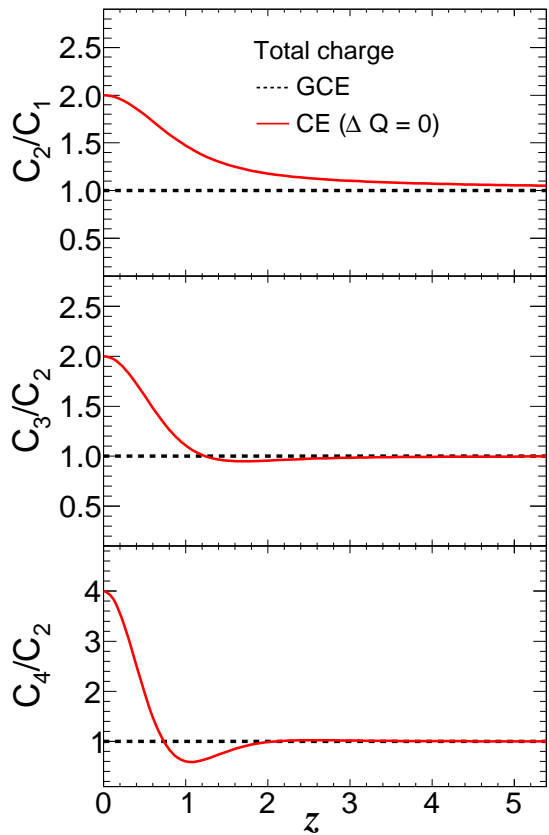}
 \caption{(Color online) Comparison of $z$ dependence of the ratio of cumulants  $C_2/C_1$, 
$C_{3}/C_{2}$ and $C_{4}/C_{2}$ for total charge in GCE (dotted line) and CE (solid 
line) for $\Delta Q$ = 0.}
 \label{fig:totch_z}
 \end{center}
 \end{figure}

\section{Results and discussion}
\label{sec:results}
The cumulants of the total charge multiplicities and their ratios are calculated in CE and GCE.
Figure~\ref{fig:totch_z} shows comparison of the ratios of cumulants for total charge 
multiplicity as a function of $z$ by considering both CE and GCE. As pointed out in 
~\cite{Begun:2004zb} that, $C_2/C_1$ calculated in GCE and CE becomes equivalent in the large 
volume limit (i.e. $z \rightarrow \infty$), it is constructive to look for the ratios of higher 
order fluctuation in two different ensembles, which might be more sensitive to the 
fluctuations. In GCE with Boltzmann approximation, the total charge multiplicities are 
strictly Poissionian, as the ratios are unity for all $z$ values, while it is not true in case 
of CE. It can be seen from figure~\ref{fig:totch_z} that $C_3/C_2$ and $C_4/C_2$ ratios in CE 
approaches to GCE for higher $z$ values. However, the cumulant ratios are quite different at 
lower $z$ values. The ratios of higher order cumulants approach the corresponding GCE values 
faster than the lower order ratio ($C_2/C_1$). 
\begin{figure}[h]
 \begin{center}
\includegraphics[width=0.95\linewidth]{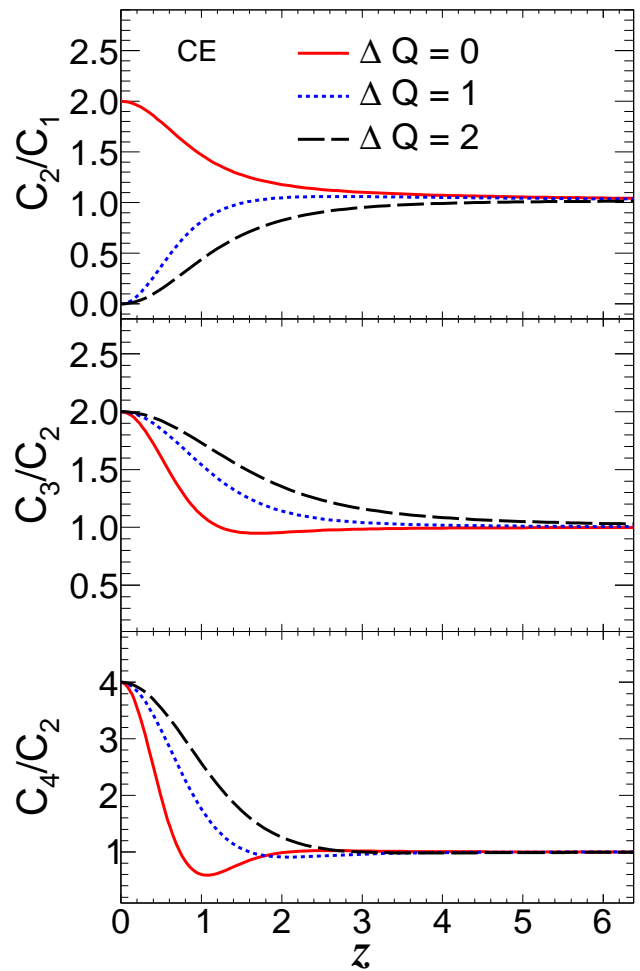}   
 \caption{(Color online) The $z$ dependence of the ratio of cumulants for total electric 
charge ($C_2/C_1$, $C_{3}/C_{2}$ and $C_{4}/C_{2}$) in canonical ensemble assuming the 
explicit net-charge of the system $\Delta Q$ = 0, 1 and 2.}
 \label{fig:zVsq}
 \end{center}
 \end{figure}
In canonical ensemble, the particle fugacity is zero in order to maintain the charge 
conservation. To study the non-zero value of conserved number fluctuation and its 
effect on different $z$ values, one can explicitly introduce the net-charge of the 
system as $\Delta Q$ = 1 and 2 as discussed in previous section. Figure~\ref{fig:zVsq}, shows 
the $z$ dependence of the ratio of cumulants for total charge with explicit net charge 
($\Delta Q$ = 0, 1 and 2) of the system. One notices that all the cumulant ratios at large $z$
(in thermodynamic limit) for different net-charge conservation approaches to 1, but the 
behavior at small $z$ is quite different. As discussed in ~\cite{Begun:2004zb}, for $\Delta Q 
\geq$ 1 the $C_2/C_1$ ratios of total charged particles decreases at smaller $z$. In case of 
small systems ($z \rightarrow$ 0), the average number of positive particles is comparable to 
the $Q$ and the fluctuations of $N_+$ are small. On the other hand, at small $z$ and fixed $Q$ 
the average number of negatively charged particles is much smaller than $Q$ and the 
fluctuations of $N_-$ are not affected by the conservation law. Hence total charge fluctuation 
is mostly driven by fluctuation of the positively charge particles.
\begin{figure}[h]
 \begin{center}
\includegraphics[scale=0.5]{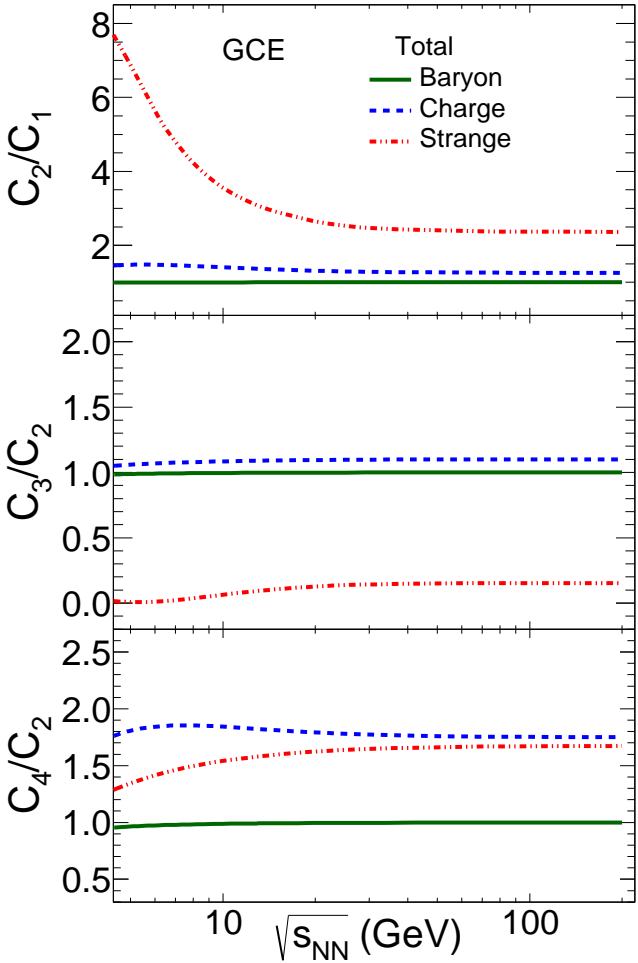}
 \caption{(Color online) The \sqsn dependence of ratio of  cumulants $C_2/C_1$, $C_{3}/C_{2}$ 
and $C_{4}/C_{2}$ for total baryons, electric charges, and strangeness fluctuations in GCE 
framework.}
 \label{fig:totmom_ene}
 \end{center}
 \end{figure} 
As can be seen, the $C_3/C_2$ and $C_4/C_2$ ratios approach 
to asymptotic value faster for $\Delta Q$ = 0 compared to non-zero $\Delta Q$ values of the 
system. Further, the ratios of higher order cumulants ($C_3/C_2$, $C_4/C_2$) approach to their 
asymptotic values at smaller $z$ values with compared to $C_2/C_1$. All the ratios of cumulants 
converges at both extremes except for $\Delta Q$ = 0 in $C_2/C_1$. 
\begin{figure}[h]
 \begin{center}
\includegraphics[scale=0.5]{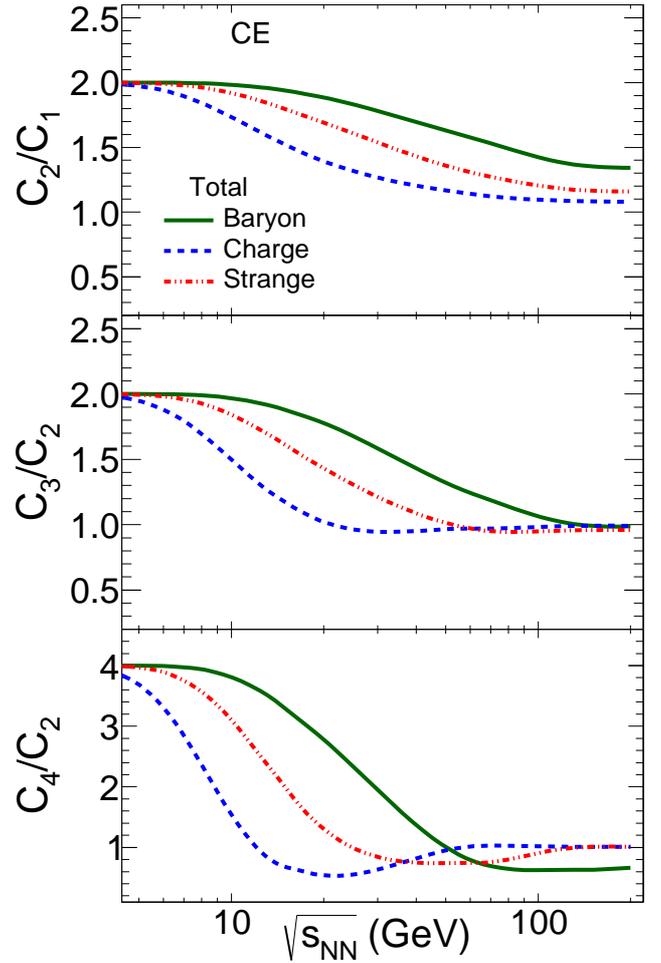}
 \caption{(Color online) The \sqsn dependence of ratio of  cumulants $C_2/C_1$, $C_{3}/C_{2}$ 
and $C_{4}/C_{2}$ for total baryons, electric charges, and strangeness fluctuations in CE 
framework assuming explicit conservation $\Delta B=\Delta Q=\Delta S$ = 0.}
 \label{fig:totmom_ce_ene}
 \end{center}
 \end{figure} 
Fluctuation of other conserved quantities such as baryon number, electric charge or 
strangeness as a function of \sqsn can be studied in the GCE and CE framework.
The cumulants of total baryon, electric charge, strangeness and their ratios are calculated in 
CE and GCE. Figure~\ref{fig:totmom_ene} shows the ratios of cumulants $C_{2}/C_{1}$, 
$C_{3}/C_{2}$, and $C_{4}/C_{2}$ in GCE as a function of collision energies \sqsn for total 
baryon, charge and strangeness calculated in thermal model approach with quantum statistics, in 
which all resonances are included to incorporate the particle interactions. The freeze-out 
parameters (baryon chemical potential $\mu_B$ and freeze-out temperature $T$) as a function of \sqsn are parametrized as ~\cite{Cleymans:2005xv}: $T(\mu_B) = a - b\mu_B^2 -c\mu_B^4$ with $a 
= 0.166 \pm 0.002 ~GeV$, $b = 0.139 \pm 0.016 ~GeV^{-1}$, and $c = 0.053 \pm 0.021 ~GeV^{-3}$. 
The energy dependence of $\mu_B$ is given as $\mu_B (\sqsn) = d/(1+e\sqsn)$ with $d$ = 1.308 
$\pm 0.028 ~GeV$ and $e = 0.273 \pm 0.008 ~GeV^{-1}$. It is to be noted that, in case of total 
baryons, the ratios of cumulants follow the Poisson expectation of individual baryons and 
anti-baryons and hence the ratios of cumulants are at unity. For heavier mass particles (when 
$m_i >> \mu$), the momentum distributions can be approximated by the classical Boltzmann 
functions, hence, the particle multiplicity in HRG model will be Poissionian. Whereas, in case 
of total charge and total strangeness, the ratios of cumulants don't follow the Poisson 
expectations in quantum statistics because of the higher charge and strangeness, $|Q|$ and 
$|S|>1$, of the particles. For total baryon and total charge all the three ratios of cumulants 
($C_{2}/C_{1}$, $C_{3}/C_{2}$, $C_{4}/C_{2}$) remain constant with collision energies. However, 
in case of total strangeness, $C_2/C_1$ decreases with increasing energies. Although CE should 
be applied to the lower collision energies where the particle multiplicities are small, we have 
carried out similar study for total multiplicities using CE. Figure~\ref{fig:totmom_ce_ene} 
shows the cumulant ratios of total multiplicity as a function of \sqsn in CE. 
At lower energies which correspond to smaller $z$ values, the cumulant ratios increase and 
higher \sqsn cumulant ratios of total baryon, charge and strangeness approach similar values. 
  \begin{figure}[ht]
  \begin{center}
    \includegraphics[width=0.95\linewidth]{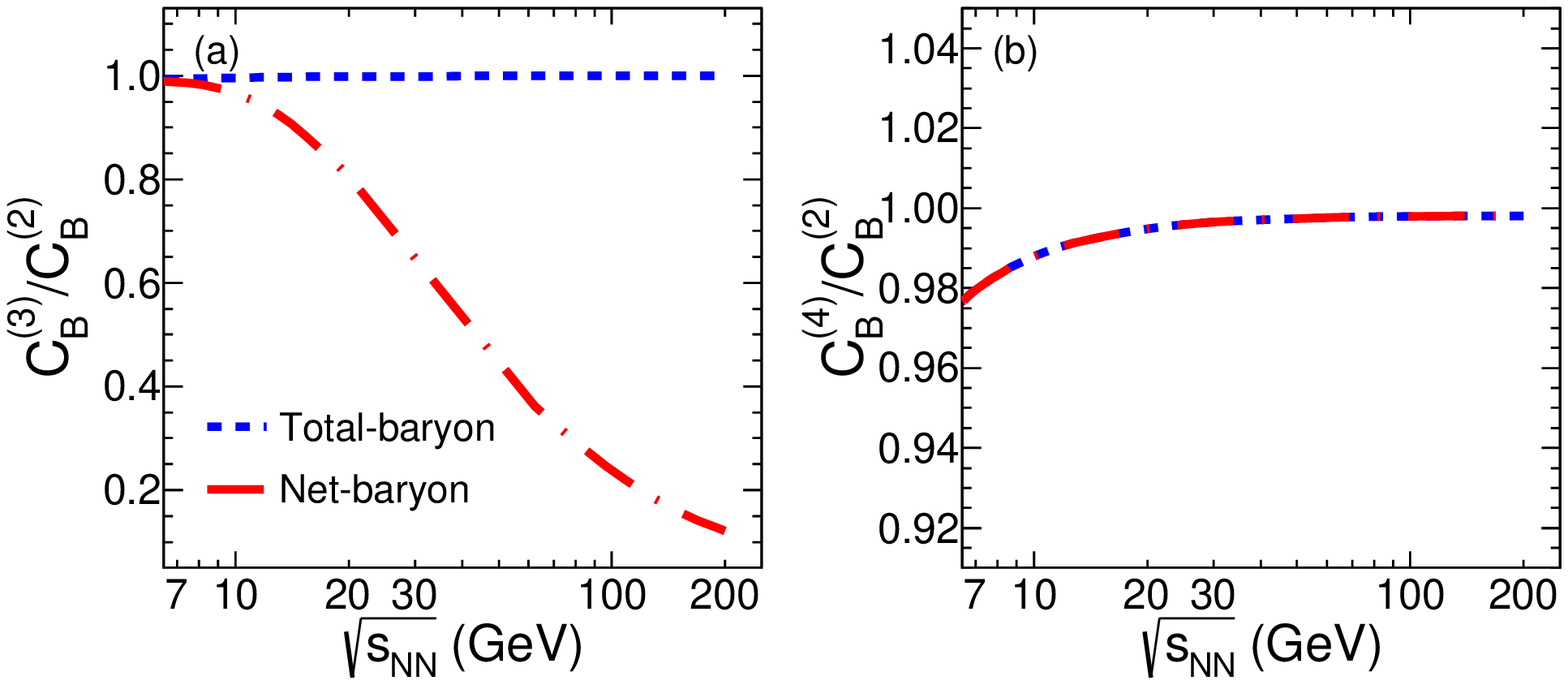}
    \includegraphics[width=0.95\linewidth]{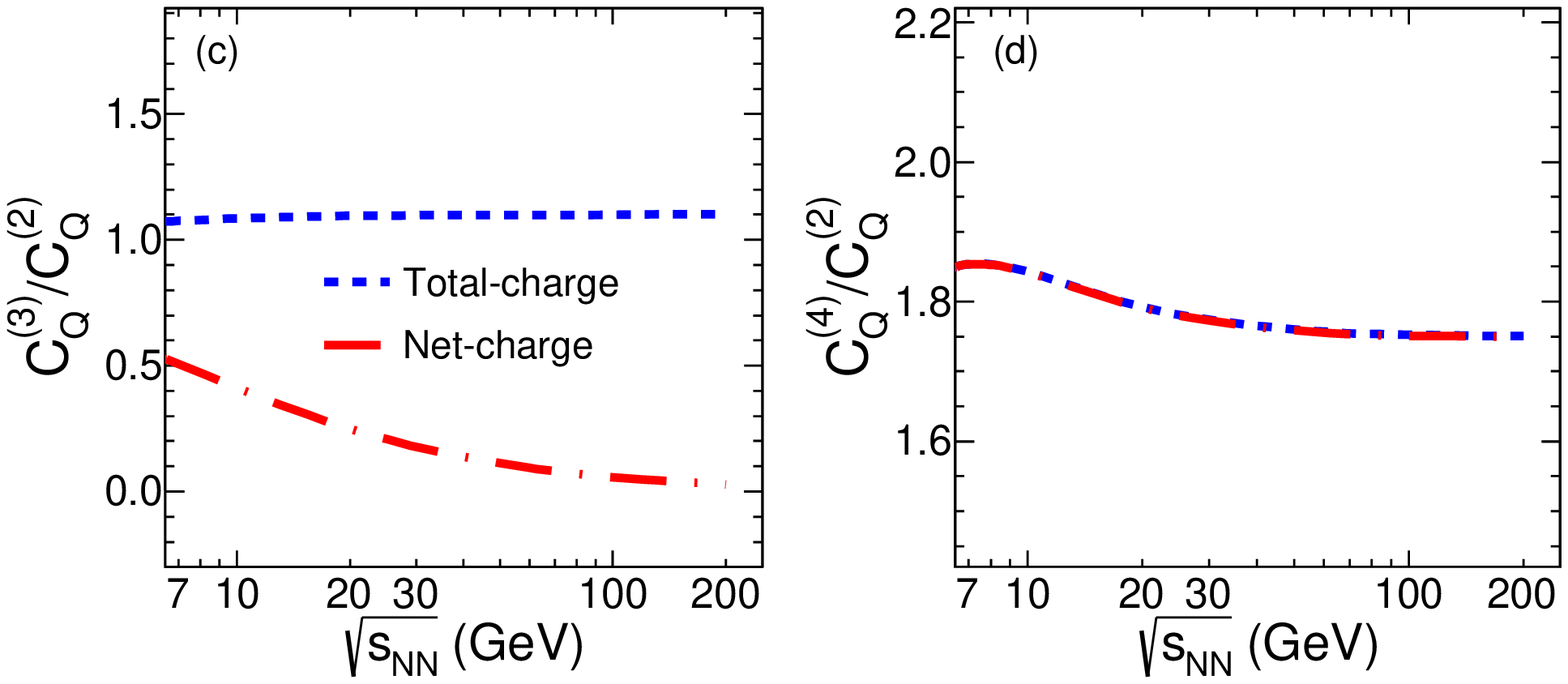}
    \includegraphics[width=0.95\linewidth]{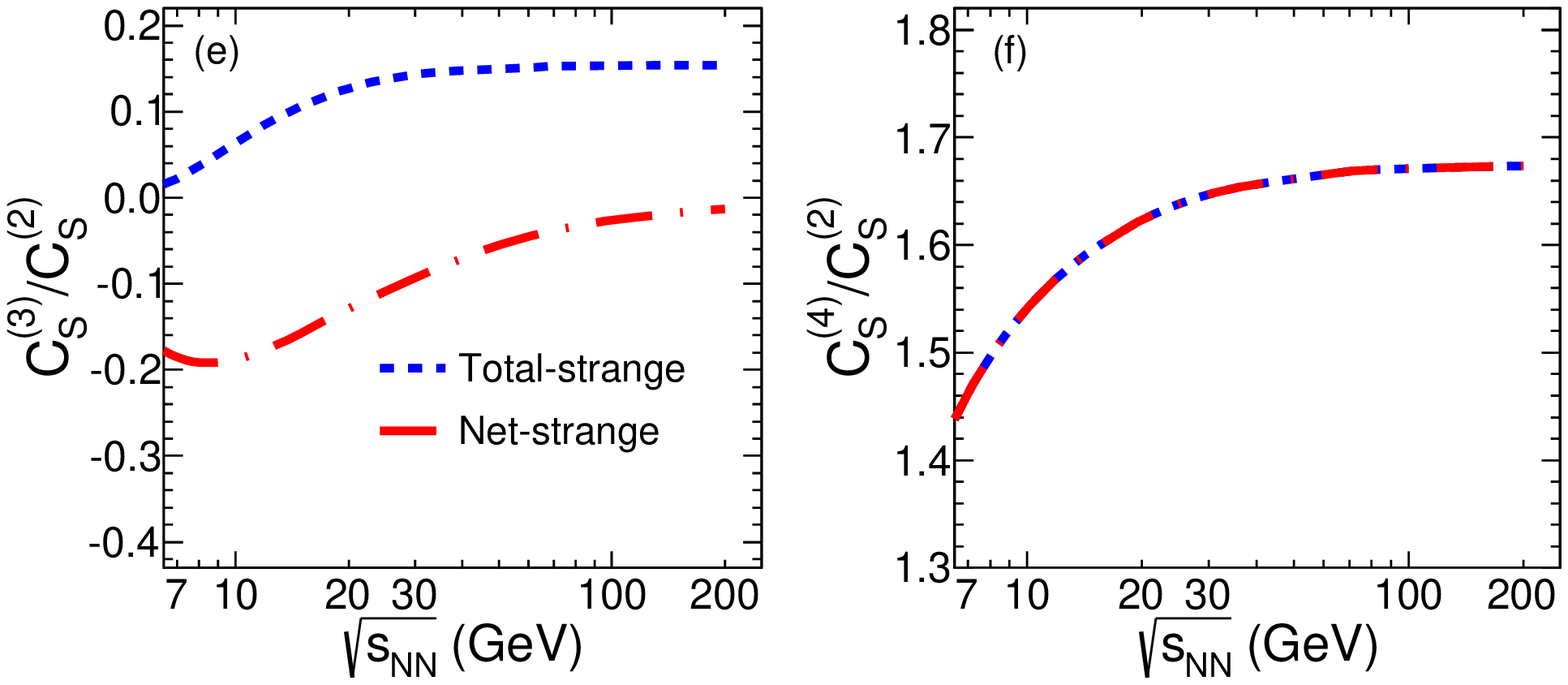}
    \caption{\label{fig:nettot}
      (Color online) The energy $\sqrt{s_{\rm {NN}}}$ dependence of $C_{x}^{(3)}/C_{x}^{(2)}$ 
and $C_{x}^{(4)}/C_{x}^{(2)}$ ratios for total (dashed) and net (dashed-dotted) conserved 
quantities using GCE. Where $x$ stands for either total baryon ($B$) (panels (a) and (b)), 
total charge ($Q$) (panels (c) and (d)), and total strangeness ($S$) (panels (e) and (f)).}
\end{center}
  \end{figure}   
\section{Comparison between total charge and net charge fluctuations in GCE}
\label{sec:totalnet}
 Recent results from RHIC BES program for net-proton, net-charge and net-strangeness 
fluctuations have been proposed to extract the freeze-out parameters and to explore the CEP in 
the QCD phase diagram \cite{Bazavov:2012vg,Karsch:2010ck}. Further, it is proposed that 
deviation of these quantities from thermal baseline would indicate the presence of CEP. At 
lower energies, anti-baryon and strangeness production is small. Hence, fluctuations such as 
net-baryons or net-strangeness are mostly dominated by proton or kaon production respectively. 
For example, net-proton fluctuations reported in~\cite{Adamczyk:2013dal,Garg:2013ata} are 
dominated by fluctuation of protons. Therefore, it is intuitive to look for the fluctuations of 
total as well as net-multiplicities of different conserved quantities. Total charge cumulants 
are calculated using Eq.~\ref{eq:cn_1}$-$\ref{eq:cn_4}, similarly one can calculate the 
cumulants for net-charge fluctuations. 
Figure~\ref{fig:nettot} shows the $C_3/C_2$ and $C_4/C_2$ ratios as a function of \sqsn
for total and net-conserved quantities. In case of net-baryon and net-charge, $C_3/C_2$ 
strongly depends on collision energies, whereas for total baryon and total charge, $C_3/C_2$ 
ratios are almost constant at all energies. For $C_4/C_2$, both total charge and net-charge 
are exactly same as a function of \sqsn. If there is no correlation between different 
particles, the various order ($n=$ 1, 2, 3 and 4) of cumulants for net-charge multiplicity can 
be written as: $C_n^{net} = C_n(N^+) + (-1)^nC_n(N^-)$, where as cumulants for total charge 
multiplicity can be written as: $C_n^{tot} = C_n(N^+) + C_n(N^-)$. The reason for this 
equivalence for $C_4/C_2$ in net-charge and total charge multiplicities is mainly because of 
cancellation of correlated terms involving particle and it's anti-particle in case of even 
order cumulants. In case of odd order cumulants the apparent differences are due to the 
contribution from correlated terms. Since higher cumulants are more sensitive to the 
fluctuation, experimentally, $C_4/C_2$ ($\sim \kappa\sigma^2$) for net-charges as a function of 
\sqsn are used to look for the non-monotonic behavior of these fluctuations which are expected 
to show large deviation from the baseline values near the CEP, if it exists. Since $C_4/C_2$ 
ratios are same for total charge and net-charge, experimentally, one should look for the ratios 
of higher cumulants of total charge distributions to look for the non-monotonic behavior as a 
function of \sqsn .

\section{Summary}
\label{sec:summary}
In summary, we have calculated the higher order cumulants and their ratios for total baryon, 
charge and strangeness multiplicity in canonical and grand canonical ensembles. 
These fluctuations in CE are further extended by explicitly introducing the net charge ($\Delta Q$ = 
0, 1 and 2) conservation, significant differences are observed for all the three cases at 
lower $z$ values. Comparing the ratios of cumulants in CE and GCE for total charge suggests 
noticeable difference for lower $z$ values. When the number of conserved quanta is small, an 
explicit treatment of these conserved charges is required, which leads to a canonical 
description of the system and the fluctuations are significantly different from the grand 
canonical ensemble. Significant differences are observed for $C_3/C_2$ ratios as a function of 
collision energies for the total charge and net-charge cases, while $C_4/C_2$ ratios are same 
in both the cases. We argue that it would be constructive to look for the fluctuations of total 
charge distributions measured experimentally for different energies and can be compared with 
the thermal baseline as discussed in the present work to look for the non-monotonic behavior. 
Further, It will be exciting to check the conserved number fluctuations at other lower energies 
of heavy ion collision data. Since the number of conserved quanta will be very small, it will 
be interesting to check whether the system follows the canonical prescription at these energies 
or not. If it follows, then fluctuations in CE should be used as a thermal model baseline to 
check any deviation due to dynamical origins. In the present work we have not incorporated various
other phenomenon for example experimental acceptance, effect of hydrodynamic flow and resonance decay.
Therefore, while comparing the experimental measurements with our calculations,
one has to take care of the above mentioned effects.

\noindent

\end{document}